# 17.   GeoAI in social science
## *Wenwen Li*

## 1.   INTRODUCTION

GeoAI, or geospatial artificial intelligence, is an exciting new area that leverages artificial intelligence (AI), geospatial big data and massive computing power to solve problems in high automation and intelligence (Li 2020; 2021). The term was first coined at an Association for Computing Machinery (ACM) workshop in 2017 and then quickly picked up by industry giants Microsoft and Esri for providing new ways of analyzing geospatial data in a cloud environment. The rapid advances of GeoAI in both academia and industry are attributed to three factors: (1) the proliferation of geospatial big data has provided abundant information for researchers to study the environment and society; (2) the recent breakthrough in AI and machine learning (especially deep learning) has better positioned AI for complex and real-world problems; and (3) the fast developments in computing technology, such as Graphics Processing Unit computing, have made it possible to run compute-intensive models using big data. GeoAI evolves as AI evolves, but it is not simply an application of AI in geography. Instead, GeoAI is an interdisciplinary field that injects spatial theories and concepts to make AI more powerful and suitable for tackling geospatial problems.

Physical science seems to be a natural adopter of the GeoAI technologies. This is in part because one prominent use of AI and deep learning is in the area of computer vision and image analysis. In physical science, there are numerous types of available Earth Observation data, such as remote sensing imagery, the analysis of which is, in essence, image analysis. Hence, there is ample room for AI and deep learning to develop in processing remote sensing images for natural and manmade features detection (Li and Hsu 2020, Konstantinidis et al. 2020), image classification and segmentation (Zhang et al. 2018), object tracking (Uzkent et al. 2017), automated surveillance (Javed and Shah 2002) and change detection (Fytsilis et al. 2016), among others. Reichstein et al. (2019) provided a comprehensive review of AI, especially relating to deep learning's applications in Earth science. Various discussions also address applications that leverage different characteristics (spatial, temporal, and spectral) of Earth science data.

The social sciences, by comparison, deal with even more diverse datasets offering both unique opportunities and challenges for the adoption of GeoAI. Nowadays, social scientists are increasingly relying on data from social media such as Twitter and Flickr to detect patterns of human movement, identify urban functions and trace breaking events such as disaster evacuations and the spread of disease (Widener and Li 2014; Hu et al. 2015). Census data collected and made available at multiple municipal levels have also been intensively used in research involving urban issues (Batty 2017), spatial econometrics (Anselin 2010) and regional inequalities (Rey et al. 2019). In the smart cities context, wireless sensor networks have been widely deployed and are being used to collect streaming data for urban traffic monitoring, analyses of indoor shopping behaviors, and tracking the flow of goods (Li et al. 2020a). Mobile phones and other Global Positioning System devices on board public transportation vehicles, taxis and share-ride bicycles are also collecting massive amounts of location information that depict human trajectories in detail (Li et al. 2020b). These datasets constitute social big data and help improve our understanding about the mechanics and dynamics of the social system.

Different from data in the physical science world – which are often image-based or composed in well-structured formats – much of the newer social science datasets, that is, social media data, remain semi-structured or unstructured. This characteristic has increased the difficulty with data collection, cleaning and preprocessing for generating analysis ready data. Issues such as noise, uncertainty and ambiguity that inherently exist in big data, especially those collected through crowd sourcing platforms, pose a significant challenge for traditional spatial analysis methods tailored for handling small, well-structured and good-quality data with less noise. In this vein, GeoAI, which has the ability to mine and learn from big data, has shown great advantages in acting at as a critical spatial analytical method to support social science applications.

The rest of the chapter is organized as follows: Section 2 describes the methodological scope of GeoAI and introduces main GeoAI techniques and their suitable application scenarios. Section 3 uses two social science use cases to demonstrate GeoAI's ability in answering questions regarding society from an unconventional perspective. Section 4 concludes the paper and discusses remaining challenges toward a seamless integration between GeoAI and social science.

## 2. GEOAI TECHNIQUES

GeoAI is a combination of AI and Geography and it evolves as AI research develops. AI, a current buzzword, is about developing machine intelligence that mimics the way humans recognize and reason about the world. Machine learning is a subset of AI. Different from other general algorithms, machine learning algorithms have the ability to learn the mapping from input data to the output results without the need for explicitly programming the analytical rules. Deep learning is a breakthrough technique of machine learning, representing the use of multiple connected neural network layers for pattern recognition in a more intelligent manner. As a new type of machine learning, deep learning techniques can automatically discern representative features from training data. Because of this, the method is very powerful in analyzing spatial processes when the mechanics of a process are not well understood.

According to the various ways of learning, GeoAI methods can be classified into top-down and bottom-up approaches. A common, top-down approach involves leveraging ontology and its recent derivative knowledge graph (Xu et al. 2016). Bottom-up approaches include the use of various learning paradigms used for clustering, classification and prediction of geospatial features or phenomena. The difference between the two classes lies in the way knowledge is represented during the machine learning and reasoning processes. In a top-down manner, the knowledge is explicitly coded in machine-understandable languages (such as Ontology Web Language) to support formal and logical reasoning. While possessing strong expressive power, the performance of a top-down approach may be affected by incompleteness of the knowledge encoded in an ontology. By comparison, bottom-up approaches are built on the premise of 'letting the data speak.' They often involve the construction of complex models and use an iterative learning process to uncover hidden patterns within the data. Most models do not need pre-defined knowledge but the insights gained are only implicitly recorded in a trained model, through for example, thousands to millions of parameters, which are often difficult to interpret. The combination of both research directions may make GeoAI even more expressive and powerful.

Further we provide a brief overview of popular GeoAI techniques from both top-down and bottom-up perspectives and discuss their suitable application scenarios in social science.

## 2.1    Bottom-Up, Data-Driven Approaches

### 2.1.1    Clustering

According to the purposes of different tasks, data-driven GeoAI methods can be classified into clustering, classification and regression. Clustering is an exploratory data analysis strategy for grouping points (or features that can be simplified into points) based on their similarity (see chapter by Helderop and Grubesic, this book). It is basically an unsupervised learning technique. Popular clustering techniques include distance-based clustering, density-based clustering and graph-based clustering. Distance-based methods measure the similarity of points based on the 'distance' between them. This measure can be geographical when the points are georeferenced, or a similarity metric may be calculated from one or more attributes of the points. A widely used method of this kind is K-means (Jain 2010). It works by assigning points to different clusters based on the distance between a point and the cluster center, which changes as the clusters are generated. K-means and its variants have been used for detecting geographical hotspots of crime, road accidents, and other social phenomena (Grubesic and Murray 2001; Jain 2010).

Density-based methods group points into clusters based on density rather than pure distance. This strategy has the advantage of clustering points (which are densely distributed) in any shape, rather than just the circular shape that is favored by the distance-based (or more precisely, the centroid-based) approach. A popular approach is DBSCAN (density-based spatial clustering of applications with noise). It grows a cluster iteratively by adding a new point if the point falls within the predefined search radius of a core point in a cluster. A core point is defined as a point that has a sufficient number $n$ of neighboring points within the given search radius $r$. Unlike K-means, this algorithm does not need a pre-defined cluster number as the input, but it needs to carefully define two other parameters ($r$ and $n$) to avoid generating too few large clusters (when $r$ is large) or labeling small clusters as noise – points which are not assigned to any cluster (when $n$ is large). DBSCAN can be used for clustering high-density social activity in urban areas (Bawa-Cavia 2011), identifying urban function zones by clustering check-in data from social media (Hu et al. 2015) and other social applications.

A third category is graph-based clustering. This clustering technique has its basis in graph theory and uses graph nodes to represent objects and links to represent strengths of connections between each node pair. The goal of graph-based clustering is to partition the graph into subgraphs (clusters) with the objective of maximizing the connections of nodes within a cluster and minimizing the connections of nodes across clusters. Therefore, it is by nature an optimization problem. Compared to distance-based and density-based clustering, graph-based clustering emphasizes connectivity among the nodes. It identifies nodes with the strongest connections and groups them together. Hence, even though two nodes are (geographically) close to each other, if the connection between them is weak, they are less likely to end up in the same cluster. The community detection algorithm is typical in this area and has been applied to the study of the spatial structure of tourism patterns (Shao et al. 2017), understanding neighborhood segregation (Prestby et al. 2019) and detecting communities in social networks (Bedi and Sharma 2016).

Clustering techniques are an important first step in analyzing and gaining insights into data, especially when patterns in the data are unknown. Besides the above-listed techniques, there are also methods such as hierarchical clustering, Gaussian Mixture Models, mean-shift clustering and so on. There are also hybrid approaches that integrate different clustering techniques to address the limitations of a single approach.

### 2.1.2   Classification and regression

AI techniques have been widely leveraged for supporting classification tasks, namely the prediction of data into different categories. If these categories involve discrete class labels, the technique is called classification. When the predicted labels are in a continuous value range, it is called regression. In general, classification is a supervised learning method, which means that ground-truth labels for some sample data are needed to train the model so that it can gain the ability to capture the mapping function between the input and output for the prediction of new observations. There are many classification techniques, such as linear techniques (that is ordinary least square [OLS]) and Geographically Weighted Regression (GWR), as well as those involving logistic regression (and naïve Bayes classifiers), support vector machines (SVMs), decision trees, boosted trees and random forests and artificial neural networks (ANNs). Each of these algorithms have a foundation in information or statistical theory.

Taking decision-tree classification as an example, this approach uses the divide-and-conquer strategy to hierarchically choose an attribute of an observation to best split the trees into homogeneous subtrees. Each tree branch (from the root to the leaf node following a certain path) is actually a decision path providing the probability of an observation belonging to some class of its attributes to satisfy criteria annotated along each tree branch. The random forest is an expansion of the decision-tree method used to avoid overfitting in using a single tree alone. It achieves this by generating multiple decision trees, each trained by a subset of the training data. The results from the decision trees are then aggregated into one final result.

Regression methods also involve supervised learning to predict the observation by a value in a continuous space. These methods, such as OLS, always have an assumption of independence among different observations. A well-known extension of OLS in spatial analysis is GWR, a localized OLS in which the influence of the sampled locations is in inverse proportion to its distance to the location for prediction (see chapter 7 by Oshan et al. in this book). Of course, there is often a debate between researchers in computer science and statistics on whether regression analysis, which usually has a strong foundation in statistical modeling, should be considered as a machine learning method. One criterion we may use to distinguish between machine learning and statistics is the purpose. Machine learning focuses more on predictive accuracy whereas statistical models, such as regression analysis, are used with the aim of characterizing the relationship among variables using mathematical equations; the prediction power is usually not its focus. When a regression model is used for prediction, it can be considered as a machine learning model. It is also worth mentioning that despite this argument, statistical learning is a foundation of machine learning and regression has been integrated into many machine learning models for classification and prediction.

There are a wide range of social science applications that can benefit from these methods, including the predicting of housing prices (Yao and Fotheringham 2016), the predicting of air quality (Kumar 2018) and the predicting of home locations based on people's social media footprints (Kavak et al. 2018).

### 2.1.3   Deep learning

All of the above models are part of the shallow machine learning family. This is because of their relatively simple model architecture, as well as the need for features engineering—a process of manually compiling a list of variables (attributes) that can impact the accuracy of predictions. Compared to these solutions, deep learning has the advantage of automatically learning prominent features important for classifying data correctly. Deep learning, or deep neural network learning, expands traditional ANN, which only contains a few layers, to a deep structure which can stack several hundred layers. In fact, the ANN architecture has existed for decades; it tries to mimic the human perception process by defining and connecting artificial neurons which can propagate information across neural network layers. A well-trained ANN can often identify complex, non-linear relationships between the output and the input, and is therefore capable of making meaningful decisions or predictions. However, the number of parameters in a fully connected ANN increases exponentially as the network goes deeper; as such the model is very difficult to converge during training because of the complexity of the model and the significant interdependency across layers.

A revolutionary breakthrough that powered AI in recent years was the introduction of convolution into an ANN architecture. Instead of stacking multiple, fully connected neural network layers, a convolutional neural network (CNN) applies convolution, a local operation between a filter sliding over the data and a subset of the data with the same size of the filter. The result is called a feature map. The CNN offers dual advantages: First, the local, convolution operation reduces computational intensity and makes the model easier to parallelize, and therefore computationally allows a network to go deep. Second, as a deep learning model, a CNN does not require features engineering. Instead, it has the ability to automatically extract prominent features from the data to make accurate predictions. Since its invention, a variety of applications have widely adopted CNN. Computer vision and pattern recognition are perhaps the areas for which CNN is the most useful, because originally it was developed to take a Red-Green-Blue (RGB) image as an input to make classifications at pixel, object and scene levels. Hence, the remote sensing community has extensively exploited CNN for mining and analyzing large amounts of available, remotely sensed images.

A sample CNN model is illustrated in Figure 17.1. It applies multiple convolution layers to the original image to gradually extract low- to high-level features, stored in feature maps. Examples of low-level features include texture, edge and shape. High-level features include the skeleton of an object and other semantic attributes. Because of its outstanding capability in learning and extracting prominent features, a CNN has become an important building block in deep learning models for accomplishing different classification tasks. For instance, after the convolution phase, a CNN can be connected with a traditional classifier such as a fully connected ANN and SVM. For image analysis, a CNN model can be integrated into a deep architecture setup to support scene classification (Krizhevsky et al. 2012), object detection (Ren et al. 2015) and semantic segmentation (Ronneberger et al. 2015).

These models, which learn to map from features to labels (class prediction results), are known as discriminative models. CNN can also be built into a model that learns in the inverse direction of a discriminative model, namely from a label (class) to representative features of data belonging to the same class. For instance, if an image has the label of 'natural scene,' a generative model can learn to generate an artificial image of a natural scene based on the features of images in this class. A Generative Adversarial Network (GAN) is a model compositing a discriminative model and a generative model with the objective of the generative model learning to create fake images that look authentic, while the discriminative model tries to make distinctions between fake and authentic images. GAN can be used to create more training samples to improve the predictive power of a classification model.

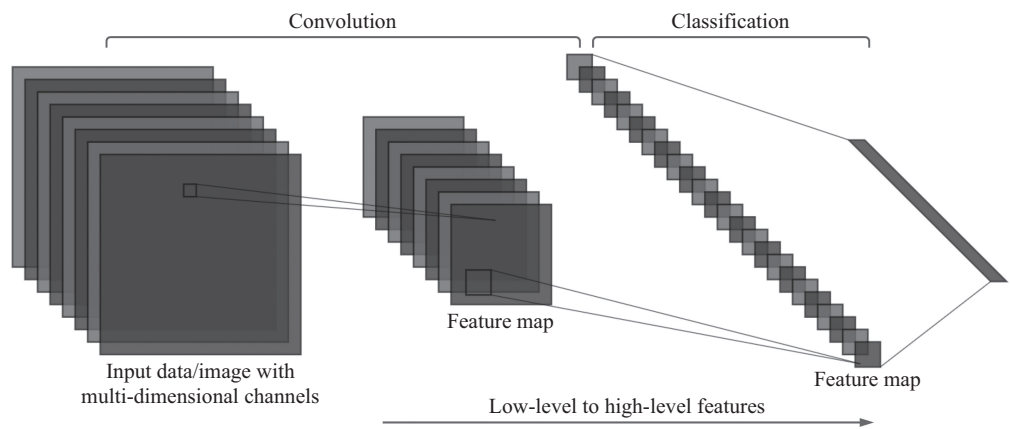

*Figure 17.1    An example of a CNN architecture*

The application of a CNN goes beyond image analysis. For multidimensional scientific data (rather than a simple RGB image), the channels of the input 'image' can be expanded to accommodate such data (Huang et al. 2016). For discovering unknown patterns in space and time, a 2D CNN can be extended to a 3D CNN in which the convolution filters become 3D so that it can capture temporal patterns in time-series data (Maturana and Scherer 2015). For 1D vector-based textual data, the CNN can be reduced into 1D CNN for text classification (Gargiulo et al. 2018). CNN can also be combined with other deep learning structures for the continual prediction of time sequence data, such as video clips, voice segments and/or text documents. For instance, researchers have leveraged a spatio-temporal LSTM (Long, Short-Term Memory) to recommend the next place for tourists to visit (Zhao et al. 2018). It works with embedding techniques (Goyal and Ferrara 2018) to learn a representation of unstructured textual data from microblogging services, news media, and other social science data sources for accurate text classification, topic modeling and geographic information retrieval (Mai et al. 2019).

## 2.2    Top-Down Ontological Approach

As discussed above, a machine learning algorithm is primarily data driven; there is usually little to no domain knowledge involved in guiding its learning process. On the contrary, an ontology-based approach tries to solve problems in a top-down manner. According to Smith (2003), 'Ontology is the science of what it is, of the kinds, structure of objects, properties, events, processes and relations in the area of reality.' It started as a philosophical term and then evolved into a computational term while keeping its philosophical meaning. Ontology has become a key supporting element for implementing the semantic network proposed in the 1960s and the Semantic Web vision in the 2000s. The objective is to build a knowledge representation framework that conceptualizes a wide range of entities, such as real-world objects, events, instances and their interrelationships. Simply put, ontology is about concepts and their relationships. Building an ontology requires input from expert knowledge through ontological engineering. Ontological objects are always hierarchically defined first, composing a tree-like structure. Nodes near the roots are more general terms and nodes near the leaves of the tree are more specific terms. Linkages can then be built among objects at the same or different levels of specificity, composing a knowledge graph (Bonatti et al. 2019).

Similar to ontology, a knowledge graph is about the formal representation of knowledge. The difference is that an ontology could be conceptual and it is usually developed by domain experts. Hence, an ontology can go very deep in the knowledge structure and used as the schema for knowledge instantiation and formalization. The focus of a knowledge graph is more on the breadth and scope of knowledge encoded in it to form a knowledge base that is machine understandable to support large-scale applications. Leading IT companies such as Google, Microsoft and Facebook all maintain an internal knowledge graph for improving their data search and developing recommendation systems.

Once an ontology or a knowledge graph is developed, semantic reasoning can be performed on top of either or both of them for dynamic knowledge discovery. For instance, given some transitive spatial properties, 'within,' according to the triple definition in the form of <subject, predicate, object>, two triples <San Diego, within, California> and <California, within, US> can be defined. Then the reasoning rules can be applied on the triples to derive <San Diego, within, US>, even without this information being explicitly defined in the ontology. Compared to machine learning, ontology development often needs to perform an abstraction of the knowledge. In addition, its representation needs to follow some formalism constraints, such as description logic. Hence, it is less flexible than machine learning because the latter can analyze data in its raw format. But ontologies and knowledge graphs can become the perfect counterparts to machine learning because they are knowledge driven and emphasize the formalization, interpretation and trustworthiness of knowledge. This characteristic is still missing in many machine learning and AI models.

In the geospatial domain, there has been pioneering work in exploiting the use of ontologies and knowledge graphs, specifically the spatially and temporally explicit knowledge graphs for achieving convergence research through advanced question answering and cyberinfrastructure technologies (Li et al. 2019; Mai et al. 2019). An ontology can be integrated into a Geographic Information System (GIS) to improve geographic information retrieval (Li et al. 2014), improve semantic understanding of land use and land cover change (Li et al. 2016) and analyze media reports and public responses during disease outbreaks (Balashankar et al. 2019), among other things. In natural language processing, it can serve as the backbone knowledge base for the extraction of a named entity (Woodward et al. 2010), placename disambiguation (Ju et al. 2016) and temporal information annotation (Neumaier and Polleres 2019). A knowledge graph is also central to knowledge preservation, providing long-term benefits for next generations. The entire geospatial knowledge discovery process, not only the prior knowledge to a study, but also the newly derived knowledge, can be encoded into the knowledge graph toward a dynamic, ever-expanding knowledge base that is machine understandable and processable.

## 3.   SOCIAL SCIENCE USE CASES

In this section, two social science use cases will be introduced that adopt the GeoAI approaches. Section 3.1 presents a novel use of deep learning and Google Street view images to understand social demography at the neighborhood level over the entire United States. Section 3.2 introduces a question answering system built upon a disease knowledge graph for the retrieval of critical events, such as location and time during a disease (that is COVID-19) outbreak.

### 3.1   Social Demography Estimation by Deep Learning and Google Street View Images

Computer scientists at Stanford University presented an interesting study using data mining, deep learning and Google Street view images to predict social demographic information such as income, educational levels and voting preferences (Gebru et al. 2017). This was perhaps one of the earliest deep learning applications in social science. Social-demographic data, which contain detailed statistics about the population, profoundly influence nearly every aspect of one's life, from finding a safe neighborhood and estimating the amount of property taxes to the planning of community and business developments. One main source for collecting such information is the American Community Survey (ACS); however, because the process is very cost and labor intensive, there is often a lag of several years between the current status and ACS statistics. To address this issue, Gebru et al. (2017) developed an automated system to infer socioeconomic variables from vehicle characteristics.

   Figure 17.2 demonstrates its data processing pipeline, which contains three stages: object detection, vehicle characteristic classification and social-demographics prediction. First, 50 million Google Street View images from 3068 zip codes and 39 286 voting precincts across 200 US cities were collected in order to retrieve local vehicle images by object detection. Next, data on 22 million automobiles were extracted which took up 32 percent of all vehicles in the 200 cities and 8 percent of all vehicles in the United States. These vehicle images were then classified into 2657 fine-grained categories consisting of combinations of car models, years, types and so on. The resulting vehicle dataset contained 88 vehicle-related features. Finally, combined with ACS and presidential election voting data, two regression models were trained to estimate demographic statistics and voter preferences from vehicle characteristics in a neighborhood. A logistic regression model was used for predicting race and educational levels; a ridge regression model was used for predicting income and voting preferences.

   The prediction results demonstrate strong correlations between the results and ACS or voting data. In city-level demographic statistics, the correlation coefficient $r$ was between 0.77 and 0.87 for race (Asians, Blacks and Whites) percentage prediction; 0.54 and 0.70 for education level prediction; and $r = 0.82$ for income prediction. For zip code-level statistics, which are more fine grained, the $r$ value for race percentage prediction was still maintained between 0.58 and 0.84. In voting preference predictions, comparing to the 2008 presidential election, the $r$ value for city-level prediction was 0.73. For precinct-level prediction, it was still highly related to the ground truth. For example, it had an 85 percent accuracy level for predicting the voting preferences of 311 precincts in Milwaukee, Wisconsin.

   Overall, this work demonstrates a nice GeoAI application in social science. By leveraging image processing, big data and deep learning, one can provide new and up-to-date demographic data at a much lower cost and with a finer-grained spatial resolution than traditional approaches, such as the ACS.

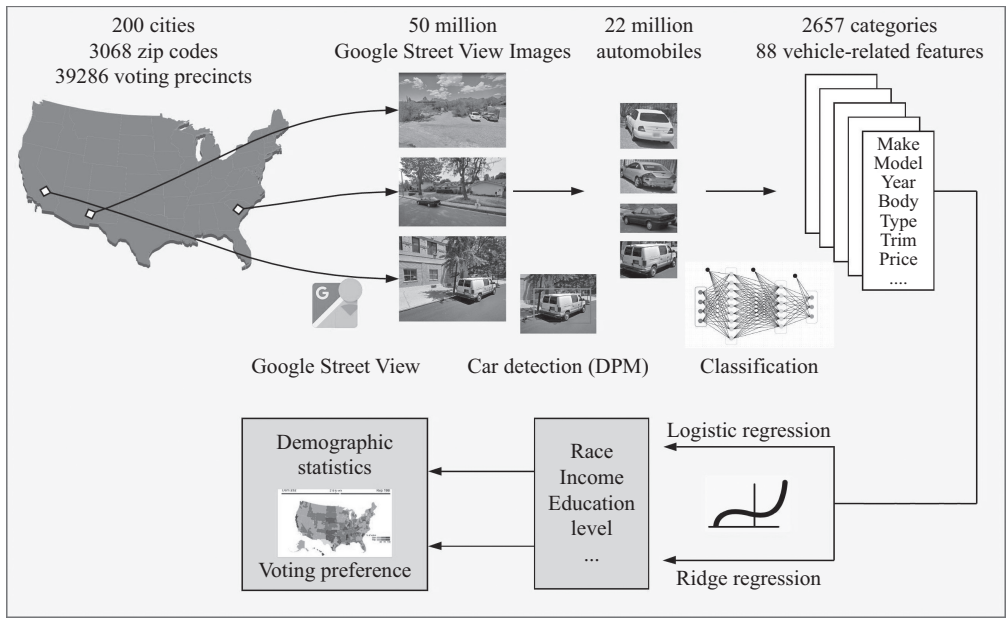

*Figure 17.2   A deep learning framework for the estimation of social demographic information*

## 3.2   A Knowledge Graph-Based Question Answering (QA) System for Tracking Disease Outbreaks

This section will introduce a use case that follows a top-down ontological approach for question answering about space, time and social events. One primary element for ensuring the performance of a deep learning model is the availability of sufficient training data. For instance, in the Section 3.1 study, more than 50 million images were collected and labeled for model training. However, in cases of rare events, such as the outbreak of a novel virus, there is often a lack of historic and current data; this limitation poses significant challenges to machine learning-based approaches. A knowledge graph, on the other hand, indexes and encodes disease-related information (that is clinical data, diagnostics, key events in the outbreak of a disease and their consequences) with fine granularity so as to provide great support to geographic information retrieval.

Consider the outbreak of COVID-19 (coronavirus disease 2019) in December 2019 in Wuhan, China, as an example. Because it was highly infectious and caused many deaths, it drew widespread attention from the media and the general public. If we were to do a Google search, although there would be many types of information returned, the sources might not provide an immediate or satisfying answer to one's question. This is due to the fact that Google search is supported by a general-purpose knowledge graph which may not include up-to-date information about knowledge in a specific domain, such as that relating to a novel disease.

Hence, a knowledge graph specifically designed for representing information about COVID-19 would offer a unique opportunity toward the building of an intelligent QA system on key information of the events related to the disease. After the disease outbreak, OpenKG.
cn, an open platform based in China, shared – for the first time – a knowledge graph covering
information about the various aspects of COVID-19. Referring to its schema and its encoded
data, we constructed a mini-COVID-19 knowledge graph to demonstrate its capability in
knowledge preservation and question answering.

Figure 17.3 shows a snapshot of the COVID-19 knowledge graph developed for tracking
important events during the disease outbreak. The definition of classes is shown in the upper
left window 'Class Hierarchy: Event;' individual elements of an event class are displayed in the
lower left window titled 'Direct instances: Event1;' general descriptions about a sample event
(*Event1*), referring to 'people-to-people spread confirmed' are displayed in the upper right
window titled 'Annotations: Event1;' and the event properties and assertions can be found in
the lower right window titled 'Property Assertions: Event1.' Once a question is given, such as,

*Figure 17.3   A snapshot of a mini-coronavirus knowledge graph and a semantic query on
top of it to support question answering*

'*When was people-to-people spread of coronavirus first confirmed in China? And by whom?*' a natural language processing engine can parse it and convert it to a formal Simple Protocol and Resource Description Framework Query Language (SPARQL) query that infers from the knowledge graph. The 'SPARQL Query' window in Figure 17.3 shows the query scripts. The bottom of that window provides a direct answer to the above question that '**ZhongNanShan**' is the person who first confirmed that the COVID-19 has shown human-to-human transmission on 20 **January 2020**. An additional answer, 'Wuhan, China,' provides the location for the event.

Besides QA, the knowledge graph will also support semantic queries. For instance, although the formal name is called 'COVID-19,' the general public may simply use 'coronavirus' in their searches. Because their synonymous relationship is explicitly encoded in the knowledge graph, the use of either keyword will return the desired answer. This query expansion mechanism also applies to the case of 'human-to-human transmission' and 'people-to-people spread.'

This case demonstrates the power of a knowledge graph in spatial and temporal information retrieval of an emergent event that has limited previous data for training the machine learning models. It is important to note that the use of a knowledge graph and machine learning are not exclusive approaches but instead complementary. For instance, machine learning is always used to help automatically construct a knowledge graph at scale and help with knowledge graph problems, such as graph summarization (Liu et al. 2018). A knowledge graph can supervise the machine learning process to make it more powerful and expressive.

## 4.   CONCLUSION

This chapter provides an overview of GeoAI, its methodology core and use cases that apply GeoAI for the social good. Although GeoAI might be relatively new to social scientists, it does offer a unique opportunity to be integrated seamlessly into social science research because of its outstanding capability in handling the diversity, volume and noise characteristics of social data that are being rapidly generated from social sensors and various observation platforms. Future research directions for achieving the convergence between GeoAI and social science are proposed below.

- Breaking the data silos

Even though the collection methods of social science data have gone beyond surveys and questionnaires and have become more automated, the data are managed by different organizations, companies and government agencies, leading to the problem of big data silos (Li 2018). For instance, transportation departments maintain the travel data of customers by public transit sources such as trains, metros, buses and so on. Telecommunications companies maintain the locations, movements, and Call Detail Records of mobile phone users. All these data provide useful information about human mobility, yet it remains difficult to depict someone's complete trajectory and obtain his or her accurate location in case of emergency (Dodge, in press). To make social big data play a greater role in emergency situations such as disease outbreaks, humanitarian crises and search and rescue operations, it is crucial for data holders to work together in developing partnerships and standards for data sharing and access. Clearly, improving data accessibility should not jeopardize user privacy. Policies and laws need to be regulated to govern the collection and proper use of social big data.

- Developing benchmark datasets for social science research

Benchmark datasets, which provide sufficient training samples, are key to ensuring the success of a machine learning algorithm. Compared to physical science domains, there have been very few benchmark social datasets available in the literature. To further deepen GeoAI applications in social science, more efforts along this line will be demanded. Strategies such as volunteered geographic information (VGI) (Goodchild 2007) and crowdsourcing can be useful for collecting such data. Projects such as YouthMappers (http://www.youthmappers.org) provide excellent opportunities for training the next generation to apply AI in mapping and at the same time contribute up-to-date spatial-social information about their local community.

- Convergence among GeoAI research methods

As discussed in Section 2, there are two mainstream GeoAI methods: bottom-up, data-driven approaches represented by machine learning and top-down, knowledge-driven approaches exemplified by ontologies and knowledge graphs. Each has its strengths and limitations. Machine learning models can gain outstanding predictive performance by relying on a deep structure and by mining from big data. However, the complexity in the models hinders interpretation of the reasoning process (Goodchild and Li 2021). On the contrary, ontologies and knowledge graphs build on a formalized definition of concepts, interrelationships and inference rules. These, therefore, have stronger expressive power but their prediction ability may not be as great as that of machine learning models. Combining the two research threads to develop a hybrid GeoAI learning framework would help address this issue (Hsu et al. 2021). That way, prior knowledge defined in the knowledge graph can be employed to supervise the machine learning process to avoid issues of overfitting and getting stuck in local optima to achieve stable and precise prediction. The use of formal knowledge and reasoning rules can also reduce AI model complexity to improve model interpretability.

- GeoAI: Moving beyond AI for geospatial benefits

Although an increasing number of studies have been reported in the area of GeoAI, the majority remain as simple imports of AI to the geospatial domain. To define the research agenda of GeoAI and to develop it as a building block research area in GIScience and beyond, it is important for us to move GeoAI research toward an in-depth fusion of geospatial sciences and AI. Explorations in how spatial principles, including spatial autocorrelation and spatial heterogeneity, can enhance AI model performance; and how spatially explicit AI models can support more accurate spatial prediction and regression (Janowicz et al. 2019; Hsu and Li 2021; Li et al. 2021; chapter by Sharma et al. in this book) will be promising research directions.

## ACKNOWLEDGEMENT

This research was in part supported by the National Science Foundation under grants 1455349, 1853864, 2021147, 2120943 and 2033521.